\documentclass[global,twocolumn]{svjour}
\usepackage{graphics}
\journalname{Applied Physics B}
\begin{document}

\title{Josephson dynamics for coupled
polariton modes under the atom-field interaction in the cavity}
\author{A.P. Alodjants\inst{1} \and S.M.
Arakelian\inst{1} \and S.N. Bagayev\inst{2} \and V.S. Egorov\inst{3}
\and A.Yu. Leksin\inst{1}} \institute{Vladimir State University,
\email{alodjants@vpti.vladimir.ru} \and Institute of Laser Physics,
Pr.Lavrentyeva 13/3, Novosibirsk, 630090 Russia Fax
(3833)33-20-67\and St-Petersburg State University}
\date{Received: date / Revised version: date}
%
\maketitle

\begin{abstract}
We consider a new approach to the problem of  Bose-Einstein
condensation (BEC) of polaritons for atom-field interaction under
the strong coupling regime in the cavity. We investigate the
dynamics of two macroscopically populated  polariton modes
corresponding to the upper and lower branch energy states coupled
via Kerr-like nonlinearity of atomic medium. We found out the
dispersion relations for new type of collective excitations in the
system under consideration. Various temporal regimes like linear
(nonlinear) Josephson transition and/or Rabi oscillations,
macroscopic quantum self-trapping (MQST) dynamics  for population
imbalance of polariton modes are predicted. We also examine the
switching properties for time-averaged population imbalance
depending on initial conditions, effective nonlinear parameter of
atomic medium and kinetic energy of low-branch polaritons.
\end{abstract}


\section{Introduction}
\label{intro}

Novel experiments on observation of Bose-Einstein condensation (BEC)
gives new opportunities to investigate of collective coherent
phenomena in quantum and atomic optics and solid state physics - see
e.g.~\cite{Keeling},~\cite{Brandes}. An important tool for that can
be taken with the help of the Josephson junction problem approach
that has a long history relating to Cooper-pair tunneling in the
superconductor Josephson junction (SJJ) systems - see
e.g.~\cite{Solymar}. Various aspects of Josephson phenomenon and
superfluid behavior for two weakly linked atomic condensates in
double-well potential have been discussed in a number of
papers~\cite{Zapata}-\cite{Meier}. However just recently the
coherent oscillations and self-trapping regime have been observed
experimentally in~\cite{Gati} for two condensates of $^{87}{\rm Rb}$
atoms placed in a special optical dipole trap. The principle item is
that photonic ``condensate'' exhibits a similar nonlinear behavior
in the problem of optical Josephson junction (OJJ) consisting of two
tunnel-coupled optical fibers and/or semiconductor waveguides
(quantum wells)~\cite{Leksin}. Another type of Josephson coupling
(nonlinear Josephson effect) for condensate modes has been
considered in~\cite{Ostrovskaya}. In particular, nonlinear dynamics
like macroscopic self-trapping and intermodal population exchange,
has been theoretically analyzed for condensate collective modes
(eigenmodes), and namely, for ground state and first excited state
modes. The effective coupling of these modes is guarantied by
atom-atom scattering, i.e. by the own nonlinearity of BEC placed in
highly anisotropic trap.

{\tolerance=1000 However, the observation of mentioned above
phenomena with pure atomic BEC medium is strictly limited by
extremely low temperature (about hundred nK) conditions. At present
the problem of high-temperature quasi-condensation phenomenon that
occurs for cavity excitons and polaritons in so-called strong
coupling regi\-me evokes a great interest~\cite{Keeling},
\cite{Deng}-\cite{Kasprzak}. Speaking more precisely the low branch
cavity polaritons can be considered as a weakly interacting
two-dimensional Bose-gas for which the Kosterlitz-Thouless phase
transition occurs at high enough temperatures due to the extremely
small mass of polaritons $m_{\rm eff}=5\cdot10^{-33}{\rm g}$ under
condition of strong coupling regime for excitons and quantized e.m.
field. Although the polariton condensate is the non-equilibrium
state and occurs in a specially control situation, the macroscopic
occupation of ground state with momentum $\vec{k} = 0$ has been
examined experimentally for quantum wells inserted in a
semiconductor (CdTe/CdMgTe) microcavity - see~\cite{LeSiDang} and
more recent results in~\cite{Richard}. The coherence properties have
been obtained in~\cite{Deng} for low-branch polaritons above the
threshold of generation in semiconductor (Ga-As) microcavity. It has
been demonstrated that statistical properties of the excitons can be
converted to match the excitons in coherent states.

}{\tolerance=50 The problem is that the BEC state in semiconductor
systems, fabricated at present, is essentially limited by disorder
factor. Recently it was suggested to use a strong disorder in the
microcavity to obtain polariton
quasi-con\-den\-sation~\cite{Malpuech}. However it is not clear
until now how it is possible to achieve true condensation for
exciton-polari\-ton substance in semiconductor microcavity for this
case.

}In our paper~\cite{Averchenko} it has been shown that polariton
true condensation can be achieved for relatively hot atomic vapor in
the case of polariton trapping, when atoms are strongly coupled with
optical field in the resonator. The degeneration of weakly
interacting 2D-gas of polaritons can be also associated with the
effect of so-called ``spectrum condensation'' observed previously
for sodium atoms in the cavity.

For some practical purposes the coherent properties of upper branch
of polaritons can be important as well. In fact, the problem of
formation of quantum entangled polariton states from different
branches~\cite{Ciuti} is the case in respect to construct the
algorithms of quantum cloning and storage for optical field into the
cavity polaritons  being  recently proposed~\cite{Alodjants}.
However, in the experiment~\cite{Kasprzak} the upper polariton
branch has not been developed in the photoluminescence when the
quasi-con\-den\-sation of lower polariton branch takes place under
the thermal equilibrium condition. The existence of large energy gap
(about ten of meV) between upper and lower branches is also a reason
for that.

{\tolerance=300 But the macroscopic population of the ground (low
branch) polaritonic state in atomic system can be only one of
possible way to obtain interbranch dynamical regimes for such a
polariton system. In particular, the energy gap in atomic system can
be in $10^{4} - 10^{5}$ times smaller then for semiconductor
one~\cite{Averchenko}, \cite{Miller}, and some (small) population of
upper branch does exist even under the quasi-equilibrium conditions.
Upper polaritonic branch can be populated also by using Landau-Zener
effect when the tunnelling phenomenon between two eigenstates of the
system takes place for the time-dependent detuning -
e.g.~\cite{Jiang}.

}Another possibility to observe such a dynamics is determined by a
special scheme of coherent excitation of polaritons from different
branches~\cite{Brunetti}. Quantum beats and Rabi oscillations in
exciton-photon system have been observed in this case~\cite{Laussy}.

{\tolerance=300 In present paper we consider the problem of
nonlinear dynamics for polaritons of different branches under the
mean field approximation. Our approach has some analogy with the
problem of macroscopic quantum tunneling and interference that takes
place for atomic condensate being associated with ground state and
first excited state. Actually, low and upper branches of polaritons
represent two types of normal modes in atom-field interaction. They
can be coupled to each other due to atomic nonlinearity.

}The paper is arranged as follows. In Section 2 we establish
nonlinear model of interaction of two-level atoms with quantized
e.m. field in the cavity for both momentum and coordinate
representation --
cf.~\cite{Keeling},~\cite{Miller},~\cite{Tartakovskii}. The
adiabatic approximation is discussed for the last case. In Section 3
we propose a new approach to describe a nonlinear Josephson junction
problem for both upper and lower macroscopic polariton modes coupled
in momentum space due to nonlinear (Kerr-like) atom-field
interaction. The analysis of Josephson dynamics and switching of
various regimes of polariton interaction is presented in Section 4.
In Conclusion we briefly summarize our results.

\section{The nonlinear model of atom-field interaction in the cavity}
\label{sec:2}

To describe the interaction of two-level atomic system with a
quantum e.m. field, we consider a localized exciton model within the
framework of the Dicke-like Hamiltonian for the interaction of the
$N$ two-level atoms with a quantized e.m. field
(cf.~~\cite{Keeling},~\cite{Brandes},~\cite{Miller}):
\begin{equation}
\label{eq1}
\begin{array}{l}
 H = \sum\limits_{\vec{k}} E_{\rm{ph}} \left(k\right) \psi_{\vec k}^{\dag}
\psi_{\vec {k}}  + \sum\limits_j^N {\frac{\displaystyle{E_{\rm{at}}}
}{\displaystyle 2}\left(b_{j}^{\dag} b_j - a_j^{\dag} a_{j}\right)}
\\ \quad \quad \quad +\sum\limits_{\vec {k}} {\sum\limits_{j}^{N} {\frac{\displaystyle{g}}{\displaystyle{\sqrt {N}}
}\left( {\psi _{\vec {k}}^{ \dag}
a_{j}^{ \dag}  b_{j} + b_{j}^{ \dag}  a_{j} \psi _{\vec {k}}}  \right)}}  \\
 \quad \quad \quad + \sum\limits_{\vec {k}} {\sum\limits_{j}^{N}
{\frac{\displaystyle{\kappa} }{\displaystyle{\sqrt {N}} }\psi _{\vec
{k}}^{ \dag} \left( {\psi _{\vec {k}}^{ \dag}  a_{j}^{ \dag}  b_{j}
+ b_{j}^{ \dag} a_{j} \psi _{\vec {k}}}  \right)}
} \psi _{\vec {k}}, \\
 \end{array}
\end{equation}
\noindent where $\psi _{\vec {k}} \left( {\psi _{\vec {k}}^{ \dag} }
\right)$ is the annihilation (creation) operator for photon with
momentum $\vec {k}$; $a\left( {a^{ \dag} } \right)$ and $b\left(
{b^{ \dag} } \right)$ are the annihilation (creation) operators for
atoms at the lower $\left| {a} \right\rangle $ and upper $\left| {b}
\right\rangle $ levels, respectively, $N$ is the total number of
particles, $E_{\rm{at}} $ is the energy of atomic transition between
the $\left| {a} \right\rangle $ and $\left| {b} \right\rangle $
levels (here we neglect the motion of atoms in the cavity),
$E_{\rm{ph}} \left( {k} \right)$ defines the dispersion relation for
photons in the cavity, and $g$ characterizes the atom-field
coupling. The last term with coefficient $\kappa $ in the
Hamiltonian (\ref{eq1}) takes into account a nonlinear atom-field
interaction, and can be considered in the framework of nonlinear
Jeans-Cummings model. The interaction implies a polariton-polariton
scattering from different branches, and is determined by Kerr-like
(optical) nonlinearity of atomic system.

The character of interbranch polariton interaction is determined by
sign of effective parameter $g\cdot\kappa N\left/E_{\rm at}\right.$.
In particular, repulsive particle interaction (with positive
scattering length) is realized when $g\cdot\kappa N<0$
($E_{\rm{at}}>0$). On the other hand, attractive interaction (with
negative scattering length) occurs for $g\cdot\kappa N>0$. In the
paper we take into account the positive coupling parameter $g$ only.

In contrast with semiconductor systems we neglect in Hamiltonian
(\ref{eq1}) by terms describing the nonlinear processes of
exciton-exciton scattering -- see e.g.~\cite{Keeling}. Such
processes can be reduced to parametric amplification of polaritons
and/or stimulated emission -- see, e.g.~\cite{Tartakovskii}. Our
approach is obviously valid for ensemble of \textit{non-} (or
\textit{weakly}) \textit{interacting} atomic gas for which the
average interatomic separation length is much larger than
characteristic length of interaction between the
atoms~\cite{Dalfovo}.

For high-reflection mirrors in the cavity the normal (to the plane
of the mirrors) component of the photon wave vector $k_{\bot}  $ is
quantized, i.e. $k_{\bot}  = \pi m \left/ L_{\rm{cav}}\right.$,
where $L_{\rm cav} $ is the cavity length, $m$ is the number of
modes. For the mode component in parallel to the mirror plane
$k_{\parallel}$ we have a continuum. In paraxial approximation
($k_{\parallel} \ll k_{\bot} $) the dispersion relation for photon
energy $E_{\rm ph}\left(k\right)$ inside the cavity can be
represented as follows:
\begin{equation}
\label{eq2} E_{\rm ph} \left(k\right) = \hbar c\left|\vec{k}\right|
= \hbar c\sqrt{k_{\bot}^2 + k_{\parallel}^2} \approx \hbar ck_{\bot}
+ \frac{\hbar^2 k_{\parallel}^2}{2m_{\rm ph}},
\end{equation}
where $m_{\rm ph}=\left.\hbar k_{\bot}\right/c$ is the effective
photon mass in the cavity.

In the paper we impose a strong coupling regime for which the cavity
modes are coherently coupled with collective atomic exitations that
can be introduced with the help of annihilation ($\phi $) and
creation ($\phi ^{ \dag} $) operators as (cf.~\cite{Alodjants}):
\begin{equation}
\label{eq3} \phi = \sum\limits_{i=1}^N \frac{a_j^{\dag}
b_j}{\sqrt{N}}\;, \quad \phi^{\dag}=\sum\limits_{i=1}^N
\frac{b_{j}^{\dag}a_j}{\sqrt{N}}\;.
\end{equation}

The low (or zero) density limit which is used for excitons in
semi-conductor structures (see e.g.~\cite{Keeling}) corresponds to
two-level atomic system without inversion. In this case the atomic
excitation operators $\phi $ and $\phi^{\dag} $ in Eqs. (\ref{eq3})
obey to usual commutation relation $\left[\phi, \phi^{\dag}\right]
\simeq 1$ for Bose system when the atoms occupy mainly a ground
state $\left| {a} \right\rangle $ of energy.

Let us consider the system (\ref{eq1})-(\ref{eq3}) of atom-field
interaction in the cavity in coordinate representation. In this case
we can introduce appropriate Lagrangian density $L$ in the form:
\begin{equation}
\label{eq4ref} \begin{array}{l} L = L_{\rm f} + L_{\rm at} + L_{\rm
int},
\\
L_{\rm f} = \frac{\displaystyle{i\hbar} }{\displaystyle{2}}\left(
{\Psi \frac{\displaystyle{\partial \Psi ^{\ast
}}}{\displaystyle{\partial t}} - \Psi ^{\ast}
\frac{\displaystyle{\partial \Psi} }{\displaystyle{\partial t}}}
\right) \\ \quad\quad+ \frac{\displaystyle{i\hbar
c}}{\displaystyle{2}}\left( {\Psi \frac{\displaystyle{\partial \Psi
^{\ast }}}{\displaystyle{\partial z}} - \Psi ^{\ast}
\frac{\displaystyle{\partial \Psi} }{\displaystyle{\partial z}}}
\right) \\ \quad\quad+ \frac{\displaystyle{\hbar
^{2}}}{\displaystyle{2m_{\rm ph}} }\left| {\nabla _{ \bot} \Psi}
\right|^{2} +
U_{\rm opt} \left| {\Psi} \right|^{2},\\
L_{\rm at} = \frac{\displaystyle{i\hbar} }{\displaystyle{2}}\left(
{\Phi \frac{\displaystyle{\partial \Phi ^{\ast
}}}{\displaystyle{\partial t}} - \Phi ^{\ast}
\frac{\displaystyle{\partial \Phi} }{\displaystyle{\partial t}}}
\right) + E_{\rm at} \left|
{\Phi}  \right|^{2},\\
L_{\rm int} = g\left( {\Psi ^{\ast} \Phi + \Psi \Phi ^{\ast} }
\right) + \kappa \left( {\Psi ^{\ast} \Phi + \Psi \Phi ^{\ast} }
\right)\left| {\Psi}  \right|^{2},
\end{array}
\end{equation}
\noindent where the $\Psi \equiv \Psi \left( {\vec {r},t} \right)$
function (i.e. the order parameter) describes both the time
evolution and spatial distribution of optical field in the cavity.
In paraxial approximation (\ref{eq2}) under the mean field theory
condition the value of $\Phi \equiv \Phi \left( {t} \right)$ is the
C-number that characterizes a macroscopic excitation in time of
atomic system (here we neglect the motion and space distribution of
atomic cloud in the cavity). In expression (\ref{eq4ref}) we also
introduce the external (harmonic) potential $U_{\rm opt} \equiv
U_{\rm opt} \left( {x,y} \right)$ for the photon trapping -
cf.~\cite{Averchenko}.

Using the expression (\ref{eq4ref}) under the variational approach
(see e.g.~\cite{Leksin}) we obtain the following coupled equations
for the field $\Psi$ and the atomic excitation $\Phi$ respectively:
\begin{equation}
\label{eq5aref} \begin{array}{l} i\hbar \left(
{\frac{\displaystyle{\partial}
}{\displaystyle{\partial t}} + c\frac{\displaystyle{\partial }}{\displaystyle{\partial z}}} \right)\Psi =\\
\quad=\left( { - \frac{\displaystyle{\hbar
^{2}}}{\displaystyle{2m_{ph} }}\left( {\frac{\displaystyle{\partial
^{2}}}{\displaystyle{\partial x^{2}}} + \frac{\displaystyle{\partial
^{2}}}{\displaystyle{\partial y^{2}}}} \right) + U_{opt} \left(
{x,y} \right)} \right)\Psi \\ \quad+ g\Phi + \kappa \left( {2\left|
{\Psi} \right|^{2}\Phi + \Phi ^{ *} \Psi ^{2}} \right)
\end{array}
\end{equation}
\begin{equation}
\label{eq5bref} i\hbar \frac{{d\Phi} }{{dt}} = E_{at} \Phi + g\Psi +
\kappa \left| {\Psi}  \right|^{2}\Psi
\end{equation}

Let us consider a solution of Eqs.~(\ref{eq5aref}),~(\ref{eq5bref})
under the \textit{adiabatic approximation} for which we suppose
$d\Phi/dt = 0$ for atomic excitation. Eliminating the $\Phi $
variable from Eq.~(\ref{eq5bref}) we arrive to self-consistent
generalized Schr\"odinger equation for the order parameter $\Psi $
that has the form:
\begin{equation}
\label{eq6ref} \begin{array}{l}i\hbar \left(
{\frac{\displaystyle{\partial} }{\displaystyle{\partial t}} +
c\frac{\displaystyle{\partial }}{\displaystyle{\partial z}}}
\right)\Psi = \left[ { - \frac{\displaystyle{\hbar
^{2}}}{\displaystyle{2m_{\rm ph} }}\left(
{\frac{\displaystyle{\partial ^{2}}}{\displaystyle{\partial x^{2}}}
+ \frac{\displaystyle{\partial
^{2}}}{\displaystyle{\partial y^{2}}}} \right)}\right.\\
\quad\left.{ + U_{\rm opt} \left( {x,y} \right) + u\left| {\Psi}
\right|^{2} + d\left| {\Psi}
\right|^{4}}\vphantom{\frac{\displaystyle{\partial^2}}{\displaystyle{\partial
t^2}}} \right]\Psi ,
\end{array}
\end{equation}
\noindent where $u = - 4{{g\kappa} }/{{E_{\rm at}} }$, $d = -
{{3\kappa ^{2}}}/{{E_{\rm at}} }$. Here we also omit the term with
${{g^{2}}}/{{E_{\rm at}} }$ that can be also eliminated with help of
simple unitary transformation.

{\hyphenpenalty=10000 \tolerance=10000 Equation (\ref{eq6ref})
describes a behavior of optical beam interacting with nonlinear
atomic medium in the cavity. The last term in Eq.~(\ref{eq6ref}) can
be discarded within the limit $\left| {d} \right|\left| {\Psi}
\right|^{2} \ll \left| {u} \right|$ that corresponds to
phenomenological (macroscopic) approach for weakly interacting
Bose-gas of photons in the cavity - cf.~\cite{Bolda}. The
coefficient $u$ that is determined by Kerr-like nonlinearity of
atomic vapor is a control parameter for that.

} {\tolerance=300 The coefficient $d$ that is always negative in
Eq.~(\ref{eq6ref}) characterizes a quintic nonlinearity of the
medium. Recently a superfluid behavior has been considered for such
a system in~\cite{PazAlonso}. In particular, formation of stable
vortex lattices for ``photonic condensate'' has been numerically
proposed even for the case without the external trapping potential
$U_{opt} \left( {x,y} \right)$ use.

}In general case when both relaxation rate of optical field $\Gamma
_{\rm f} $ and atomic excitations dephasing rate $\Gamma _{\rm ex} $
are under consideration the adiabatic approach is valid when an
inequality
\begin{equation}
\label{eq7ref} \Gamma_{\rm ex} \gg \Gamma_{\rm f}
\end{equation}
\noindent is hold - cf.~\cite{Lax}.

In the present paper we consider the atom-field interaction in the
limit when the rates $\Gamma_{\rm ex} $ and $\Gamma_{\rm f} $ are
the same order of magnitude. In this case we can examine the system
described by Hamiltonian~(\ref{eq1}) using in fact the two
macroscopically populated polaritonic modes from different brunches.

\section{Boson Josephson junction (BJJ) model for polaritons in the cavity}
\label{sec:3}

Let us consider a Josephson dynamics for two modes, i.e. upper and
low branch polaritons, those are coherently excited in the cavity --
cf.~\cite{Laussy}. We suppose that one quantum optical mode (with
$\vec{k} = 0$ as a limit) is only macroscopically occupied. Thus, in
Hamiltonian (\ref{eq1}) we switch over  to macroscopic (continuous)
variable $\psi \left(\vec{k}\right)$ for optical field describing
this mode. In this case we introduce the \textit{time-dependent}
classical wave functions\linebreak $\Phi _{1,2} \left( {t} \right)
\equiv \sqrt {N_{1,2} \left( {t} \right)} e^{i\theta_{1,2} \left(
{t} \right)}$ as:
\begin{eqnarray}
\label{eq8aref}
\Phi_1&=&\frac{1}{\sqrt{2}}\left(\psi(\vec{k})-\phi\right),\\
\label{eq8bref}
\Phi_2&=&\frac{1}{\sqrt{2}}\left(\psi(\vec{k})+\phi\right),
\end{eqnarray}
that characterizes the bosonic quasi-particles in atomic medium
corresponding to upper and lower branch polaritons; $N_{1,2}
\left(t\right)$ is the time dependent average number of polaritons
for upper ($N_1$) and lower ($N_2$) branches respectively,
$\theta_1$ and $\theta_2$ are the phases.

Now we introduce the new variables $\varsigma$ and $\theta$ as
\begin{equation}
\label{eq9} \varsigma = \frac{N_1 - N_2}{N_{\rm ex}}, \quad \theta =
\theta_2 - \theta_1,
\end{equation}
\noindent that characterize a fractional state population imbalance
($\varsigma$) and relative phase ($\theta$) for polaritonic modes
respectively, $N_{\rm ex} = N_1 + N_2$ is the total number of
polaritons in the cavity.

In the paper we pay our attention to the problem of polariton
interaction along the cavity length under the atom-field resonance
condition $\Delta \equiv E_{\rm at}-\hbar c \left|
k_{\bot}\right|=0$. The Hamiltonian (\ref{eq1}), describing the
superfluid dynamics of the polaritons in momentum representation,
can be evaluated in the terms of new variables $\varsigma$ and
$\theta$ as
\begin{equation}
\label{eq10}
\begin{array}{l}
H = \Lambda_{\rm eff} \beta \varsigma^2 - \Lambda_{\rm eff} \beta
\left(\sqrt{1 - \varsigma^2} \cos\theta - \sqrt{1 + \beta^2}\right)
\\ \quad \quad \times \sqrt{1 - \varsigma^2} \cos\theta
+ \Delta E_{\rm eff} \varsigma,
\end{array}
\end{equation}
\noindent where $\beta = \left. E_{\rm tr}\right/ |g|$ is the scaled
(normalized) energy of polaritons in transverse plane, $E_{\rm tr} =
\left. \hbar^2 k_{\parallel}^2\right/\left(2m_{\rm pol}\right)$ is
their kinetic energy producing the effect of quantum pressure,
$m_{\rm pol} \simeq 2m_{\rm ph}$ is the mass of polaritons in the
cavity. It is possible to define Josephson-like length $\lambda_{\rm
J}=\sqrt{|\beta|}\left/k_{\parallel}\right.$ that specifies a
spatial control parameter in the system -- cf.~\cite{Solymar},
\cite{Raghavan}. For strong coupling regime with the value of
atom-field coupling parameter $g \simeq 0.05$~meV the length
$\lambda_{\rm J} \simeq 11$~$\mu{\rm m}$. The result is the same as
for atomic BEC in the BJJ problem -- cf.~\cite{Raghavan}. The
crucial dimensionless parameter $\Lambda_{\rm eff} = \left.\kappa
N_{\rm ex}\right/\left[2|g|\left(1 + \beta^2\right)\right] \equiv
\Lambda\left/\left(1 + \beta^2\right)\right.$ determines a nonlinear
Josephson coupling between polaritons of different branches (the
parameter $\Lambda<0$, i.e. $\kappa N_{\rm{ex}}<0$, for positive
scattering length of polaritons).

In expression (\ref{eq10}) for Hamiltonian $H$ we define an
effective zero-point energy difference $\Delta E_{\rm eff} $ as
\begin{equation}
\label{eq11} \Delta E_{\rm eff} = \alpha - \Lambda_{\rm eff} \sqrt
{1-\varsigma^2} \left(1-\beta^2\right)\cos\theta,
\end{equation}
\noindent {\tolerance=300 where we have made a denotation $\alpha =
\sqrt{1 + \beta^2} \left(1 + \Lambda_{\rm eff}\right)$. Physically,
the parameter $\alpha$ characterizes the energy gap between
polariton branches, i.e. determines the unalterable zero point
energy difference when condition $\beta=1$ takes place.

}Using the Hamiltonian (\ref{eq1}) and definition (\ref{eq11}) for
$\Delta E_{\rm eff} $ one can obtain the following equations for
canonically conjugated variables $\varsigma$ and $\theta$:
\begin{equation}
\label{eq12} \begin{array}{l} \frac{\displaystyle
d\varsigma}{\displaystyle d\tau} = - \Lambda_{\rm eff}
\left(\varsigma \left(1 - \beta^2\right) - \beta \sqrt{1 + \beta^2}
\right. \\ \quad \quad \left. + 2\beta \sqrt{1 - \varsigma^2}
\cos\theta \right)\sqrt{1 - \varsigma^2} \sin\theta,
\end{array}
\end{equation}
\begin{equation}
\label{eq13}
\begin{array}{l}
\frac{\displaystyle d\theta}{\displaystyle d\tau} = \alpha -
\Lambda_{\rm eff} \left(\frac{\displaystyle 1 -
2\varsigma^2}{\displaystyle \sqrt{1 - \varsigma^2}}
\right)\cos\theta \\ \quad \quad + 2\Lambda_{\rm eff} \beta \left(
\varsigma - \frac{\displaystyle \sqrt{1 + \beta^2}
\varsigma}{\displaystyle 2\sqrt{1 -
\varsigma^2}}\cos\theta + \varsigma \cos^2\theta \right. \\
\quad \quad \left. + \frac{\displaystyle \beta}{\displaystyle
2}\left(\frac{\displaystyle 1 - 2\varsigma^2}{\displaystyle
\sqrt{1-\varsigma^2}}\right)\cos\theta \right),
\end{array}
\end{equation}
\noindent where $\tau = gt/\hbar$ is the dimensionless time
coordinate.

Thus, the set of equations (\ref{eq12}), (\ref{eq13}) represents an
extension of the BJJ model in nonlinear dynamics approach with the
phenomenon of tunneling for polaritons from different branches. In
the case $\beta=0$ the Eqs.~(\ref{eq12}),~(\ref{eq13}) describe the
polariton arising effect along the cavity length only.

Obviously, it is possible to establish some link between the system
of Eqs.~(\ref{eq12}), (\ref{eq13}) describing the polariton dynamics
in the cavity and the other Josephson dynamical systems --
cf.~\cite{Solymar}~--~\cite{Ostrovskaya}. In particular, the
equation (\ref{eq12}) characterizes the interbranch  current $I$ for
polaritons. In general it has a complex form. However we can
represent the current $I$ for small values of $\beta$ as
\begin{equation}
\label{eq14}
\begin{array}{l}
I \simeq - I_{\rm c} \sqrt{1 - \varsigma^2} \sin\theta +
J_1 \varsigma \sqrt{1 - \varsigma^2} \sin\theta \\
\quad \quad \quad + I_{\rm c} \left(1 - \varsigma^2\right)\sin
2\theta,
\end{array}
\end{equation}
\noindent where $I_{\rm c} = \left.{{|\kappa| N_{\rm ex} E_{\rm tr}}
}\right/\left({{2\left| {g} \right|\hbar}}\right)$ is the positive
critical current for polaritons in transverse plane (we took
$\Lambda_{\rm{eff}} < 0$), $J_1 = \left.|\kappa| N_{\rm
ex}\right/\left({{2\hbar}}\right)$ is the current characterizing the
polaritons along the cavity length.

In general the interbranch current $I$ is essentially nonlinear one;
the case is quite different from the Cooper-pair current in the SJJ
problem -- cf.~\cite{Solymar}, \cite{Zapata}. From Eq. (\ref{eq14})
it is easy to see that the first term, describing the supercurrent
of polaritons in transverse plane, is completely similar to the BJJ
inertrap current occurring for atomic condensates. The last term in
Eq. (\ref{eq14}) represents a nondissipative current determined by
interaction of the polaritons inside the cavity. Physically, this
term can be associated with the second order contribution to
Josephson current due to the effect of tunneling for atoms in
condensate state to noncondensate quasiparticles in the BJJ
problem~\cite{Meier}. For small population imbalance $|\varsigma|
\ll 1$  the contribution to the current of polaritons along the
cavity length can be neglected .

Energy difference $\Delta E_{\rm eff}$ defined in Eq.~(\ref{eq11})
plays a principal role in total polariton dynamics. In respect of
the atomic BJJ problem the zero-point energy difference $\Delta
E_{\rm eff} $ is determined by both the geometry of the trap and the
nonlinearity of the atomic condensates~\cite{Raghavan},
\cite{Zhang}. The energy $\Delta E_{\rm eff} $, determined by
applied voltage across the junctions, results in oscillations being
in analogy with the ac-Josephson frequency $\omega _{\rm ac} \sim
\Delta E_{\rm eff} $ in the SJJ problem. However in the case of
polaritons, the energy difference $\Delta E_{\rm eff} $ in Eq.
(\ref{eq11}) is mostly depends on dynamical variables $\varsigma $
and $\theta$ in general. Physically, such a behavior is caused by
nonlinear (Kerr-like) cross-interaction of polaritons from different
branches in the cavity. The energy difference $\Delta E_{\rm eff}$
defined in Eq. (\ref{eq11}) can be suppressed for the case
$\Lambda_{\rm eff}<0$. For example, we have $\Delta E_{\rm eff} = 0$
for the values $\beta=1$, $\Lambda _{\rm eff} = - 1$ what implies
the polaritons with momentum $k_{\parallel}\simeq
1\left/\lambda_{\rm J}\right.$ and kinetic energy $E_{\rm tr} =
\left.|\kappa| N_{\rm ex}\right/4$ . This condition is compatible
with the requirement for definition of the healing (or coherent)
length $\xi_{\rm c} $ that plays an important role for superfluid
systems - cf.~\cite{Dalfovo}. In particular, a typical length scale
$\xi_{\rm c}$ for that is determined by the condition when the
$E_{\rm tr}$ energy is of the order of the interaction energy for
condensate atoms (determined by the $\kappa N_{\rm ex} $ parameter).
For the case under discussion we have the coherent length $\xi_{\rm
c} \sim \lambda_{\rm J} $.

\section{Nonlinear superfluid dynamics for trapped polaritons}
\label{sec:4}
\subsection{Small-amplitude oscillations}
\label{sec:41}

We start our analysis from the case of \textit{non-interacting
(ideal) gas} of polaritons when $\Lambda = 0$. The solutions of
Eqs.(\ref{eq12}), (\ref{eq13}) are simply presented in the form:
\begin{equation}
\label{eq15} \theta = g\sqrt{1+\beta^2} t \equiv \frac{\Delta
E}{2}t, \quad \varsigma = \varsigma(0),
\end{equation}
\noindent where $\Delta E = 2g\sqrt{1+\beta^2} $ is the energy
splitting for non-interacting polaritons, $\varsigma(0)$ is the
initial population difference. Thus, in the case under consideration
the polariton branches are not overlapped and therefore the
polaritons arising in the problem are still uncoupled.

Now let us consider a \textit{small-amplitude oscillations} case for
population imbalance $\varsigma$ when the time average values
$\left\langle\theta\right\rangle = 0$ (``zero phase'' oscillations)
or $\left\langle\theta\right\rangle = \pi $ (``$\pi $-phase''
oscillations). Linearizing the Eqs.~(\ref{eq12}), (\ref{eq13}) in
respect of the $\varsigma$ and $\theta$ variables we arrive to
equation
\begin{equation}
\label{eq16} \frac{d^2\varsigma}{d\tau^2} + \Omega_{0,\pi}^2
\varsigma = F_{0,\pi}  \quad ,
\end{equation}
\noindent where two parameters, i.e. dimensionless angular
frequencies of oscillations $\Omega_{0,\pi}$ and  external forces
$F_{0,\pi}$, are defined as
\begin{equation}
\label{eq17a} \Omega_{0,\pi}=\left(\Omega_{\rm JP\;
0,\pi}^2\pm\Omega_{\rm R\; 0,\pi}^2\right)^{1/2},
\end{equation}
\begin{equation}
\label{eq17b} \Omega_{{\rm JP}\;0,\pi}=\sqrt{E_{\rm J}^{0,\pi}E_{\rm
C}^{0,\pi}},
\end{equation}
\begin{equation}\label{eq17c}
\Omega_{{\rm R}\;0,\pi}  = \sqrt {\Lambda_{\rm eff} \left( {1 -
\beta^2} \right)\left( {\alpha \mp \Lambda_{\rm eff} \pm
\Lambda_{\rm eff} \beta^2} \right)},
\end{equation}
\begin{equation}\label{eq17d}
F_{0,\pi}  = - E_{\rm J}^{(0,\pi)}\left(\alpha\mp\Lambda_{\rm
eff}\pm\Lambda_{\rm eff}\beta^{2}\right).
\end{equation}
\noindent In expressions (\ref{eq17a})--(\ref{eq17d}) we introduce
two characteristic (dimensionless) parameters, i.e. the Josephson
coupling energy $E_{\rm J}^{(0,\pi)} = \Lambda_{\rm eff} \beta
\left(2 \mp \sqrt{1 + \beta^2}\right) \simeq I_{\rm c}$, and
``capacitive'' energy $E_{\rm C}^{(0,\pi)}=\Lambda_{\rm eff} \beta
\left(4 \mp \sqrt {1 + \beta^{2}}\right)$ that describe the
polariton nonlinear interaction for different branches --
cf.~\cite{Gati}.

The expressions (\ref{eq17a})--(\ref{eq17c}) determine the
dispersion relations for a new type of collective excitations
arising in the cavity for polariton interaction (we stress here the
fact that $\beta$-parameter is proportional to momentum as $\beta
\sim k_{\parallel}^2$). Nevertheless it is useful to establish some
link between the problem considered here and some other Jo\-sephson
junction systems. In particular, the small amplitude oscillation
frequency regime, described by Eq.~(\ref{eq17a}), has the same
structure as a frequency of small oscillations for the atomic
population imbalance occurring for Bose-Einstein condensate confined
in the W-potential~\cite{Raghavan}. The second term in Eq.
(\ref{eq17a}) characterizes the linearized Rabi oscillations with
frequency $\Omega_{\rm R\;0,\pi}$ of the population imbalance. The
angular frequency $\Omega_{\rm JP}$ in Eqs.~(\ref{eq17a}),
(\ref{eq17b}) can be considered as an analog of the plasma frequency
for Josephson plasmon in the problem of a Cooper pair macroscopic
tunneling process across the junction -- cf.~\cite{Solymar},
\cite{Paraoanu}. The driving force $F_{0,\pi}$ depends on
$\beta$-parameter that determines quantum pressure of polaritons in
transverse spatial plane.

We distinguish two principal regimes of collective amplitude
oscillations for the cavity polaritons. Namely, within the limit
$\Omega_{\rm R} \gg \Omega_{\rm JP}$ so called Rabi regime occurs.
For quasiparticles with a small momentum $k_{\parallel}$ (i.e. for
$\beta^{4} \ll 1$) the population imbalance oscillates with angular
frequency $\Omega _{0} \approx \sqrt{4.5\Lambda_{\rm
eff}^2\beta^2-0.5\Lambda_{\rm eff}\beta^2+\Lambda_{\rm eff}}$ vs
$\tau$ under the zero phase regime. In the case of absence of
polaritons in transverse plane ($\beta=0$) we have $\Omega_0 =
\Omega_{\rm R\;0} = \sqrt{\Lambda}$ (for $\Lambda>0$). The
expression (\ref{eq17b}) and numerical simulation show that the
Josephson coupling energy $E_J$ is vanished, and the contribution of
Rabi frequency $\Omega_{\rm R} $ in Eq.~(\ref{eq17a}) is important
for parameter $\beta\to1.7$.

On the other hand, the Josephson regime is realized for the
polaritons in the cavity within the limit $\Omega_{\rm R} \ll
\Omega_{\rm JP}$. Such a regime of polariton oscillations is
achieved for the parameter $\beta = 1$.

{\tolerance=1000 In the limit of large $\beta$ ($\beta^2 \gg 1$),
i.e. for the quasiparti\-cles with large momentum $k_{\parallel}$,
the frequency of am\-plitude oscillations is
$\Omega_0\simeq\sqrt{\Lambda_{\rm eff}^2\beta^4 + \left|\Lambda
_{\rm eff}\right|\beta^3} = \sqrt {\sqrt{\left|\Lambda_{\rm
eff}\right|\beta}E_{\rm C} \Omega_{\rm R} + \Omega_{\rm R}^2}$ (for
$\Lambda_{\rm eff}<0$) that is in good agreement with the
oscillation frequency expression obtained for atomic population
imbalance in the BJJ problem~\cite{Paraoanu}. Thus, cross\-over to
Rabi oscillations with frequency $\Omega_{\rm R} \simeq \sqrt
{\left| {\Lambda _{\rm eff}} \right|\beta^3}$ (for $\left| {\Lambda
_{\rm eff}} \right|\beta \ll 1$) can be easily achieved by varying,
for example, of the $\Lambda$-parameter (i.e. the intensity of laser
field).

}The solution of Eq.~(\ref{eq16}) is given by following expression:
\begin{equation}
\label{eq18} \varsigma_{0,\pi}  = C\cos\left(\Omega_{0,\pi}
\tau\right) + \varsigma_{\rm dis}^{\left(0,\pi\right)},
\end{equation}
\noindent where $C$ is the constant determined by initial
conditions, $\left\langle {\varsigma _{0,\pi}  \left( {\tau}
\right)} \right\rangle = \varsigma_{\rm dis}^{\left( {0,\pi}
\right)} = F_{0,\pi}\left/\Omega_{0,\pi}^2\right.$ characterizes the
displacement for the time average fractional population imbalance
$\left\langle\varsigma\right\rangle $ in respect of the zero value.
The displacement $\varsigma_{\rm dis}^{\left(0,\pi\right)}=0$ and
the driving forces $F_{j} = 0$ for $\beta = 0$ or $\beta = 1$,
$\Lambda_{\rm eff} = - 1$. In this case small amplitude oscillations
of the $\varsigma$ variable occur around the value
$\left\langle\varsigma\right\rangle = 0$.

Developed above approach can be violated, for example, for
running-phase regime when, first, the phase difference $\theta$
grows unrestrictedly and, second, the essentially nonlinear
oscillations of population imbalance $\varsigma$ take place. We are
also emphasis here some difficulties to obtain so-called Fock regime
by using the Hamiltonian (\ref{eq1}) when fulfillment of the
inequality $E_{\rm c} \gg E_{\rm J} $ is required --
cf.~\cite{Paraoanu}.

\subsection{Stationary states of polaritons}
\label{sec:42}

First, we focus our attention on the \textit{out-of-phase}
stationary states of polaritons in the cavity determined by the
conditions $\sin\theta \ne 0$, $\cos\theta \ne 0$. From
(\ref{eq12}), (\ref{eq13}) it is easy to obtain the following
solutions
\begin{equation}
\label{eq20}
\begin{array}{l}
\displaystyle\varsigma_0 =
-\frac{\displaystyle\beta}{\displaystyle\Lambda}
\frac{\Lambda+2}{\sqrt{\beta^2+1}}, \\ \theta_0 =
\arccos\frac{\displaystyle{\Lambda - \beta^2 + 1}}{\displaystyle{\rm
sgn}\left(\Lambda\right)\sqrt{\Lambda^2 -
4\beta^2\left(\Lambda+1\right)}},\end{array}
\end{equation}
\noindent that represents unstable saddle point with energy $H_0 = -
\frac{\beta}{\Lambda}\left(\Lambda + 1\right)$. The existence of
solutions (\ref{eq20}) implies fulfillment of conditions for
population imbalance $\varsigma$ and phase $\theta$, i.e.
$\left|\varsigma\right| \le 1$, $\left|\cos\theta\right| < 1$. The
fact results in certain condition for parameters $\beta$ and
$\Lambda$ that looks like:
\begin{equation}
\label{eq21} \Lambda < - \frac{1}{2}\left(\beta^2 + 1\right)
\end{equation}

Second, we consider \textit{another out-of-phase} solution of
Eqs.~(\ref{eq12}), (\ref{eq13}) with $\sin\theta = \pm 1$. For
stationary fractional population imbalance $\varsigma$ and energy
$H$ of the system we find:
\begin{equation}
\label{eq22}
\begin{array}{c}
\varsigma_{1,2} = \pm 1, \\ H_{1,2} = \pm \sqrt {\beta^2 + 1} \pm
\frac{\displaystyle{\beta^2 + 1}}{\displaystyle{ \pm 2\beta +
\sqrt{\beta^2 + 1}} }\left(1 \pm
\frac{\displaystyle{\beta}}{\displaystyle{\sqrt{\beta^2 + 1}}
}\right),\end{array}
\end{equation}
\begin{equation}
\label{eq23}
\begin{array}{c}
\varsigma_3 = \frac{\displaystyle{\beta \sqrt {\beta^2 + 1}}
}{\displaystyle{1 - \beta^2}} = - \frac{\displaystyle{\left(
{\Lambda + \beta^2 + 1} \right)\sqrt {\beta^2 + 1} }} {\displaystyle{2\Lambda \beta} }, \\
H_3 = \frac{\displaystyle{\beta^3}}{\displaystyle{1 -
\beta^2}}.\end{array}
\end{equation}

The solutions (\ref{eq22}), (\ref{eq23}) do exist if the
parameters $\beta$ and $\Lambda$ fulfill the conditions :
\begin{equation}
\label{eq23_2} \Lambda = \frac{\displaystyle\left(\beta^2 +
1\right)^{3/2}}{\displaystyle\pm 2\beta + \sqrt{\beta^2 + 1}}\quad
\rm{for} \quad \varsigma_{1,2},
\end{equation}
\noindent and
\begin{equation}
\label{eq24_2} \Lambda = \beta^2 - 1 \quad \rm{for} \quad
\varsigma_3,
\end{equation}
\noindent where $\beta \in \left[-
\frac{1}{\sqrt{3}};\frac{1}{\sqrt{3}}\right]$. Thus, the solutions
(\ref{eq22}), (\ref{eq23}) represent the unstable saddle points in
phase portrait.

Third, Eqs.~(\ref{eq12}), (\ref{eq13}) have \textit{zero-phase
(in-phase) and} $\pi$\textit{-phase} stationary solutions with the
phase $\theta$ determined by the condition $\sin\theta = 0$. The
stationary points in this case can be found out from solution of
algebraic equation
\begin{equation}
\label{eq24} A_1\varsigma^4 + A_2\varsigma^3 + A_3\varsigma^2 +
A_4\varsigma + A_5 = 0
\end{equation}
\noindent for fractional population imbalance $\varsigma$, where the
real coefficients $A_j \equiv A_j\left(\Lambda,\beta\right)$ are the
functions of $\Lambda$ and $\beta$ (we are not directly represent
them in the paper due to complexity of the relevant expressions). In
general, there exist four roots of Eq.~(\ref{eq24}) of which only
the real solutions are of interest. Note, that one of the solutions
of (\ref{eq24}) may become a saddle point if the condition
(\ref{eq21}) is not fulfilled.

\subsection{Phase plane portrait description} \label{sec:43}

From mathematical point of view the equations (\ref{eq12}),
(\ref{eq13}) describe an undamped nonrigid pendulum with phase
$\theta$ and canonical momentum $p_\theta  \equiv \varsigma$. In the
paper we consider a phase-space description to understand the
dynamics of the polariton interaction in the cavity. In Fig.1 the
evolution of the phase portraits for various values of normalized
kinetic energy $\beta$ of polaritons is presented for the same
nonlinear $\Lambda$-parameter. The trajectories are plotted for
different values of initial conditions for the population imbalance
and the phase with $\Lambda$ and $\beta$ parameters kept constant.
Closed loop trajectories in the gray area marked in Fig.1 correspond
to \textit{periodic oscillations} of pendulum coordinate and phase
in terms of nonlinear pendulum analogy. The stationary points $S_j$
($j=1,...,6$) correspond to the roots of Eq.(\ref{eq24}). White
color regions 1 in Fig.1 determine the \textit{running phase}
\textit{regimes} that correspond to anharmonic rotation of nonrigid
pendulum with relative phase $ - \infty < \theta < + \infty$.
Stationary saddle points $P_{1,2}$ at the separatrixes in Fig.1a,b
are determined from expressions (\ref{eq20}) respectively.

\begin{figure}
\resizebox{0.455\textwidth}{!}{%
  \includegraphics{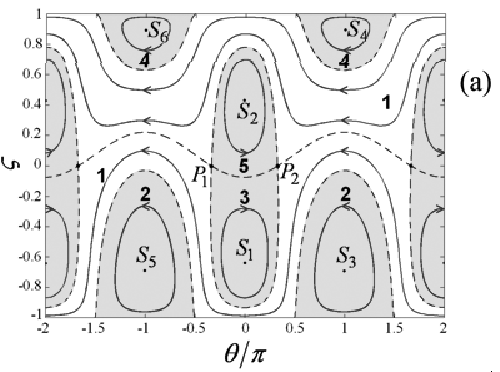}
}
\resizebox{0.455\textwidth}{!}{%
  \includegraphics{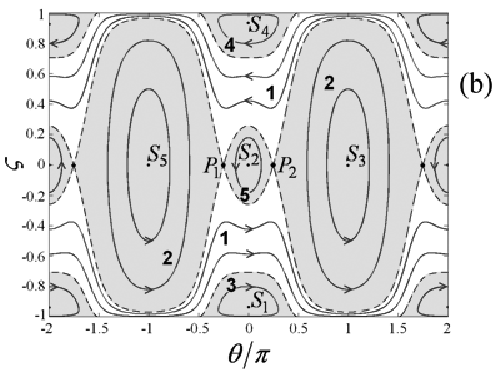}
}
\resizebox{0.455\textwidth}{!}{%
  \includegraphics{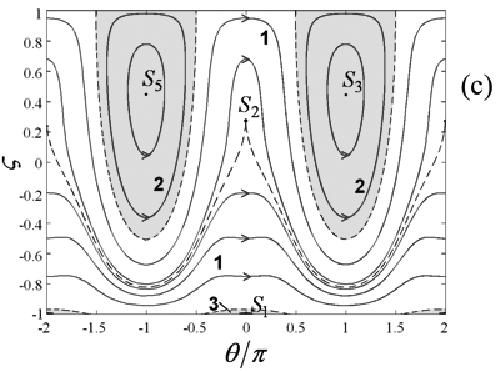}
}
\caption{Phase-plane portraits of the variables $\varsigma$ and
$\theta /\pi$ for: a) $\beta = 0.15$, b) $\beta = 1$, c) $\beta =
2.08475$. For all cases the parameter $\Lambda = - 2$. The points
$S_j$ and $P_j$ are the stationary points. The regions numerated as
`1,2,\ldots,5' determine the various dynamical regimes separated by
dashed curves (separatrixes).}
\label{fig:2}       
\end{figure}

For arbitrary value of zero-point energy difference $\Delta E_{\rm
eff}$ defined in (\ref{eq11}) the phase-plane portraits are
asymmetric as a rule. However, the phase portrait established in
Fig.1b becomes completely symmetric for $\Delta E_{\rm eff} = 0$. In
particular, the small amplitude \textit{zero phase Josephson
oscillations} of fractional population imbalance take place here in
the region 5 with the frequency $\Omega_0 = \Omega_{\rm JP} $ around
the average value $\left\langle {\varsigma} \right\rangle = 0$ with
zero displacement. The fact is in agreement with
Eqs.~(\ref{eq17a})--(\ref{eq18}). Two symmetric regions numerated as
2 in Fig.1b characterize the large-amplitude ($\pi$-phase)
oscillations of pendulum around the top of vertical axis. A complete
energy transfer between polaritons from different branches takes
place in this case approximately. For small initial values of the
fractional population imbalance $\varsigma^2(0) \ll 1$ we obtain the
small-amplitude $\pi$-phase\textit{ }oscillations with frequency
$\Omega_{\pi}$ -- see (\ref{eq17a}).

Other regions in Fig.1b and all of them in Fig.1a,c specify
so-called {\it macroscopic quantum self-trapping regi\-mes (MQST)}
for which a fractional population imbalance $\varsigma$ oscillates
around the \textit{non-zero} value $\left\langle {\varsigma}
\right\rangle \ne 0$. It implies that population of either of
polariton branches is never completely depleted. In fact, in our
case a macroscopic quantum self-trapping behavior of polaritonic
mo\-des takes place, first, as a result of nonlinear interaction of
macroscopically large number of polaritons from different branches.
Second, the MQST regime occurs as a result of the ``external force''
action (see (\ref{eq16}), (\ref{eq18})) depending on effective
difference $\Delta E_{\rm eff} \ne 0$ in (\ref{eq11}).

In first case, the dependence on normalized time $\tau$ for
fractional population imbalance $\varsigma$ is presented in Fig.2 to
visualize the various MQST regimes. In fact, the MQST effect at the
phase $\theta = 0$ occurs for several cases: in two regions 3,5 in
Fig.1a, in regions 3,4 in Fig.1b, and in region 3 in Fig.1c. The
curve 1 in Fig.2 demonstrates a self-trapping behavior of polariton
interaction for last case. The regimes discussed here appear for
\textit{localized (periodically) phase}. A similar dynamic MQST
behavior occurs for nonlinear interaction of two atomic BEC modes
(fundamental and first excited) under the coupled two-mode
approximation -- cf.~\cite{Ostrovskaya}.
\begin{figure}
\resizebox{0.43\textwidth}{!}{%
  \includegraphics{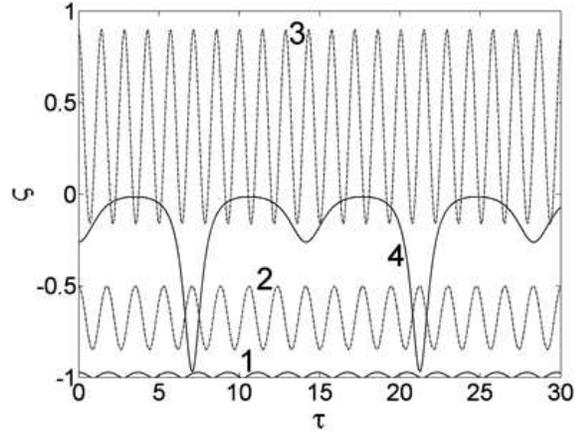}
}
\caption{Fractional population imbalance $\varsigma$ against scaled
dimensionless time $\tau$. The parameters are: $\beta = 2.08475$,
$\varsigma(0) = -0.97$, $\theta(0) = 0$ for curve 1, $\beta = 0.15$,
$\varsigma(0) = -0.5$, $\theta(0) = \pi$ for curve 2, $\beta =
2.08475$, $\varsigma(0) = 0.9$, $\theta(0) = \pi$ for curve 3 and
$\beta = 1$, $\varsigma(0) = -0.259$, $\theta(0) = 0$ for curve 4.
For all cases the parameter $\Lambda=-2$.}
\label{fig:3}       
\end{figure}

The second kind of the MQST regime associated with so called
$\pi$\textit{-phase MQST mode} is presented in Fig.1 for relative
phase values $\theta = \pm \pi$. The gray-color regions marked as 2,
4 in Fig.1a and as 2 in Fig.1c are responsible for that and
characterize the $\pi$-phase pendulum rotations with closed loop
trajectories around the value $\left\langle {\varsigma}
\right\rangle \ne 0$ -- cf.~\cite{Raghavan}. In Fig.2 the curves 2,3
demonstrate the MQST dynamics discussed in respect of the regions
taken from the Fig.1a and Fig.1c. The curve 3 describes the
\textit{large amplitude} MQST oscillations appearing here for the
time-averaged population imbalance $\left\langle {\varsigma}
\right\rangle \ne 0$ displacement as a result of presence of
zero-point energy difference $\Delta E_{\rm eff} \ne 0$.

We are emphasis here the two types of the $\pi$-phase modes
occurring in the MQST regions in Fig.1a,c for the phase values
$\theta = \pm \pi$. Two various dynamical trapping regimes (above
and beyond the stationary $\varsigma$-symmetry breaking value of
population imbalance $\varsigma_{\rm s}$) have been firstly
discussed for macroscopic atomic trapping in the BJJ
problem~\cite{Raghavan}. The modified conditions for that can be
represented as follows (cf.~\cite{Zhang}):
\begin{equation}
\label{eq25} \left\langle {\varsigma}  \right\rangle < \varsigma(0),
\end{equation}
\noindent for I type of MQST, and
\begin{equation}
\label{eq26} \left\langle {\varsigma}  \right\rangle > \varsigma(0)
\end{equation}
\noindent for II type of MQST. At the stationary (threshold) point
$\varsigma_{\rm s} $ we have $\left\langle\varsigma\right\rangle =
\varsigma(0) = \varsigma_{\rm s} $, i.e. the average population
imbalance coincides with it initial value $\varsigma(0)$.

{\tolerance=500 In contrast with the BJJ problem in our case for
atomic condensate $\pi$-phase modes one can obtain a new type of
\textit{zero-phase} MQST modes when the system demonstrates the
self-trapping effect according to the conditions (\ref{eq25}),
(\ref{eq26}). In fact, we speak about the MQST states appearing in
the regions 3,5 (Fig.1a), in regions 3,4 (Fig. 1b) and in region 3
(Fig.1c). For stationary points $S_{1,4}$ in Fig.1b we have
$\left\langle {\varsigma} \right\rangle = \varsigma_{\rm s}$.

}Finally, third kind of the MQST dynamical regime occurs for various
\textit{running phase modes} with unbounded phase $\theta$. The case
is presented by white-color regions 1 in Fig.1. The curve 4 in Fig.2
characterizes the MQST appearing in the region 1 in Fig.1b.

\subsection{Switching dynamics for polaritons}
\label{sec:44}

The MQST regimes under consideration can be achieved by different
ways depending on the crucial parameters $\beta$, $\Lambda$ and also
on the values of initial conditions for relative phase $\theta(0)$
and fractional population imbalance $\varsigma(0)$. Practically we
can vary one of them keeping others unchanged. For example, let us
consider transition from the regime 5 of the Rabi oscillations in
Fig.1b upward, i.e. to the MQST regime 1 with unbounded running
phase. We can establish the condition for such a transition in this
particular case in the form:
\begin{equation}
\label{eq27} H\left( {\varsigma \left(0\right),\theta
\left(0\right)} \right) > H_{\rm sep}\;,
\end{equation}
\noindent where $H\left(\varsigma(0),\theta(0)\right)$ is the
initial (conserved) energy of the system defined in (\ref{eq10}),
the energy $H_{\rm sep} = - \frac{\beta}{\Lambda}\left(1 +
\Lambda\right)$ characterizes the state that corresponds to two
separatrixes going through the points $P_{1,2}$ in Fig.1b. For
instance, in the case when the parameters $\Lambda$, $\beta$ and
initial phase $\theta(0)$ are unchanged the bound energy $H_{\rm
sep} = 0.5$ determines a critical value of fractional relative
population imbalance $\varsigma_{\rm c} \approx 0.2588$. The MQST
dynamics occurs in the case when initial value $\varsigma(0) >
\varsigma_{\rm c}$.

In Fig.3a we plot the dependence for time-averaged population
imbalance $\left\langle {\varsigma}  \right\rangle$ as a function on
ratio $\varsigma(0)\left/\varsigma_{\rm c}\right.$. The value
$\left\langle {\varsigma} \right\rangle = 0$ in Fig.3a characterizes
a polariton interaction for the region 5 in Fig.1b. Sharp changing
of population imbalance at $\varsigma \left( {0} \right) =
\varsigma_{\rm c}$ implies the ``phase transition'' to the MQST
state with $\left\langle {\varsigma} \right\rangle \ne 0$ (region
1). By similar manner one can obtain a switching effect for
dynamical regimes by varying of the $\Lambda$-parameter.

\begin{figure}
\resizebox{0.46\textwidth}{!}{%
  \includegraphics{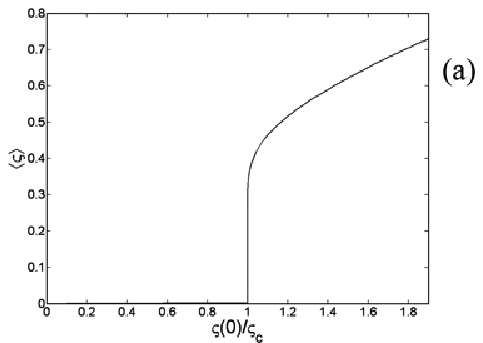}
}
\resizebox{0.46\textwidth}{!}{%
  \includegraphics{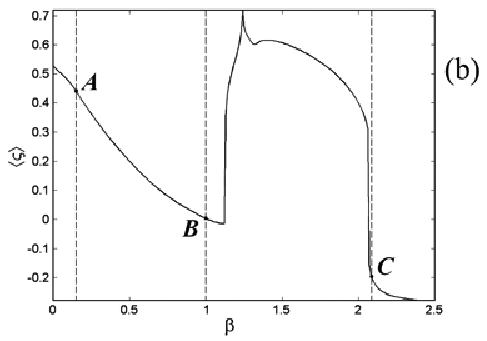}
}
\caption{Time-averaged population imbalance
$\left\langle\varsigma\right\rangle$ versus (a) --
$\varsigma\left(0\right)\left/\varsigma_{\rm c}\right.$ and (b) --
$\beta$-parameter. The parameters for (a) are: $\varsigma_{\rm
c}\simeq 0.2588$, $\beta = 1$, and for (b) --
$\varsigma\left(0\right) = 0.2$. In all cases parameter $\Lambda = -
2$ and initial phase $\theta\left(0\right) = 0$.}
\label{fig:4}       
\end{figure}

In Fig.3b we establish a switching phenomenon for the time-averaged
population imbalance $\left\langle {\varsigma}  \right\rangle$
depending on $\beta$-para\-meter (i.e. on kinetic energy of
polaritons in transverse plane). The points A,B,C mark the values of
$\beta$-parameter for the phase portraits presented in Fig.1.
Initial value of $\left\langle {\varsigma} \right\rangle$ at
$\beta=0$ characterizes the average fractional population
im\-balance for polaritons along the cavity length when the
su\-per\-current $I_{\rm c} = 0$ -- see Eq.~(\ref{eq14}). The curve
AB determi\-nes the transition from the MQST regions in Fig.1a to
the region 5 of the Josephson oscillations in Fig.1b. Then, with
increasing the absolute value of $\beta$-parameter, the regions 5
and 4 are unified in phase plane and form a new drop-like region
where the system is self-trapped (is not shown in Fig.1). Such an
evolution of the phase portrait in phase-plane is accompanied by
sharp transition presented in Fig.3b near the value of $\beta
\approx 1.2$. Therefore, the drop-like region is collapsed to
stationary point $S_2$ in Fig.1c with further expanding of the
$\beta$-pa\-ra\-me\-ter. The curve BC in Fig.3b describes a
switching effect from the MQST state with locked phase to the
running phase MQST states determined by the region 1 in Fig.1c.

The switching dynamics of polaritons depending on the
$\beta$-parameter has an important feature in the framework of
polariton condensation problem for atomic system. In particular, for
initially non-equilibrium state of polaritons in the cavity with
large momentum $k_{\parallel}$, the variation of $\Lambda$-parameter
can be used in experiment to achieve a quasi-condensation with
$k_{\parallel}=0$ for low-branch polaritons. In turn, the last
parameter can be easily varied in the experiment by changing the
total number $N_{\rm ex}$ of polaritons in the cavity (we can
change, for example, the intensity of laser field interacting with
the atoms). Thus, observed dynamical regimes for both fractional
population imbalance $\varsigma$ and relative phase $\theta$ allow,
first, to estimate the fraction of condensed polaritons in the
cavity taking into account the relevant expression (\ref{eq14}) for
the current $I_{\rm c}$ and, second, to manipulate with those
coherence properties.

\section{Conclusion}
\label{concl}

{\tolerance=200 In the paper we developed a quantum theory of
Josephson dynamics for two macroscopic polariton modes (belonging to
upper and lower energy branches being coupled together due to
Kerr-like nonlinearity of atomic medium). The phenomenon can be used
to examine the macroscopic superfluid properties of condensed
polariton state as well. The analogy with nonrigid pendulum behavior
has been carried out to understand the different regimes of
polariton interaction. We found out the dispersion relations and the
characteristic frequencies for various regimes of collective
excitations  in the cavity including the several effects, i.e. the
Josephson and Rabi oscillations, different kinds of the MQST states
for fractional relative population imbalance $\varsigma$. Some of
them can be represented as the new one, and  have not been
considered yet.

}{\tolerance=400 The realization of macroscopic tunneling and
interbranch interaction for polariton modes is important to study in
experiment the relative phase coherence in the presence of polariton
condensation (and/or quasi-con\-den\-sation).

}
\begin{acknowledgement} This work was supported by the Russian
Foundation for Basic Research (projects Nos. 04-02-17359,
05-02-16576) and the Federal Programs of the Ministry of Science and
Education of Russian Federation. Authors are very grateful to
Referees for the valuable comments. A.P.A. is grateful to Mikhail
Glazov for fruitful discussions.
\end{acknowledgement}


\begin{thebibliography}{}
\bibitem{Keeling}
J. Keeling, F.~M. Marchetti, M. H. Szymanska, P. B. Littlewood:
Semicond. Sci. Technol. \textbf{22,} R1 (2007).
\bibitem{Brandes}
T. Brandes: Phys. Rep. \textbf{408/5-6,} 315 (2005).
\bibitem{Solymar} L. Solymar, \textit{Superconductive
Tunneling and Applications} (Chapman and Hall, London 1972).
\bibitem{Zapata}
I. Zapata, F. Sols, A.~J. Leggett: Phys. Rev. A \textbf{57,} R28
(1998).
\bibitem{Raghavan}
S. Raghavan, A. Smerzi, S. Fantoni, S.~R. Shenoy: Phys. Rev. A
\textbf{59,} 620 (1999).
\bibitem{Zhang}
Y.-B.Zhang, H.~J.~W. Muller-Kirsten: Eur. Phys. J. D \textbf{17,}
351 (2001).
\bibitem{Meier}
F. Meier, W. Zwerger: Phys.Rev. A \textbf{64,} 033610 (2001).
\bibitem{Gati}
R. Gati, M. Albeiz, J. Folling, B. Hemmerling, M.~K.~Oberthaler:
Applied Phys. B \textbf{82,} 207 (2006).
\bibitem{Leksin}
A.~Yu. Leksin, A.~P. Alodjants, S.~M. Arakelian: Optics and Spectr.
\textbf{94,} 768 (2003).
\bibitem{Ostrovskaya}
E.~A. Ostrovskaya, Y. Kivshar, M. Lisak, B. Hall, F.~Cattani, D.
Anderson: Phys. Rev. A \textbf{61,} 031601 (2000).
\bibitem{Deng}
H. Deng, G. Weihs, C. Santori, J. Bloch, Y. Yamamoto: Science
\textbf{298,} 199 (2002).
\bibitem{LeSiDang}
Le Si Dang, D. Heger, R. Andre, F. Boeuf, R. Romestain: Phys. Rev.
Lett. \textbf{81,} 3920 (1998).
\bibitem{Richard}
M. Richard, J. Kasprzak, R. Andre, R. Romestain, Le Si Dang, G.
Malpuech, A. Kavokin: Phys. Rev. B \textbf{72,} 201301(R) (2005).
\bibitem{Kasprzak}
J. Kasprzak, M. Richard, S. Kundermann, A. Baas, P. Jeambrun, J. M.
J. Keeling, F. M. Marchetti, M. H. Szymanska, R. Andre, J. L.
Staehli, V. Savona, P. B. Littlewood, B. Deveaud, Le Si Dang: Nature
\textbf{443,} 409 (2006).
\bibitem{Malpuech}
G. Malpuech, D. Solnyshkov, H. Ouerdane, M. Glazov, I. Shelykh:
Phys. Rev. Lett. \textbf{98,} 206402 (2007).
\bibitem{Averchenko}
V.~A. Averchenko, A.~P. Alodjants, S.~M. Arakelian, S.~N.~Bagayev,
E.~A. Vinogradov, E.~S. Egorov, A.~I.~Sto\-lyarov, I.~A.~Chekhonin:
Quantum Electronics \textbf{36,} 532 (2006).
\bibitem{Ciuti}
C. Ciuti: Phys. Rev. B \textbf{69,} 245304 (2004).
\bibitem{Alodjants}
A.~P. Alodjants, S.~M. Arakelian, S.~N. Bagayev,
I.~A.~Che\-kho\-nin, E.~S. Egorov: J. of Rus. Laser Research
\textbf{27,} 400 (2006).
\bibitem{Miller}
R. Miller, T.~E. Northup, K.~M. Birnbaum, A. Bocca, A.~D. Boozer,
H.~J. Kimble: J.Phys.B: At. and Mol. Opt. Phys \textbf{38,} S551
(2005).
\bibitem{Akulin}
V. M. Akulin, W. P. Schleich: Phys. Rev. A \textbf{46,} 4110 (1992).
\bibitem{Jiang}
S. Jiang, S. Machida, Y. Takiguchi, Y. Yamamoto, H. Cao: Appl. Phys.
Lett. \textbf{73,} 3031 (1998).
\bibitem{Brunetti}
A. Brunetti, M. Vladimirova, D. Scalbert, M. Nawrocki, A. V.
Kavokin, I. A. Shelykh, J. Bloch: Phys. Rev. B \textbf{74,} 241101R
(2006).
\bibitem{Laussy}
F. Laussy, M. Glazov, A. Kavokin, D. Whittaker, G. Malpuech: Phys.
Rev. B \textbf{73,} 115343 (2006).
\bibitem{Tartakovskii}
A.~I. Tartakovskii, D.~N. Krizhanovskii, D.~A. Kurysh, V.~D.
Kulakovskii, M.~S. Skolnick, J.~S. Roberts: Phys. Rev. B
\textbf{65,} 081308(R) (2002).
\bibitem{Dalfovo}
F. Dalfovo, S. Giorgini, L.~P. Pitaevskii, G. Stringari: Rev. Mod.
Phys. \textbf{71,} 463 (1999).
\bibitem{Bolda}
E. L. Bolda, R. Y. Chiao, W. H. Zurek: Phys. Rev.Letts \textbf{86,}
416 (2001).
\bibitem{PazAlonso}
M. J. Paz-Alonso, H. Michinel: Phys. Rev. Letts. \textbf{94,} 093901
(2005).
\bibitem{Lax}
M. Lax, \textit{Fluctuation and coherence phenomena in classical and
quantum physics} (Gordon and Breach, New York 1969).
\bibitem{Paraoanu}
G.-S. Paraoanu, S. Kohler, F.Sols, A.~J. Leggett: J.Phys.B: At. and
Mol. Opt. Phys \textbf{34,} 4689 (2001).
\end{thebibliography}
\end{document}